\documentclass[superscriptaddress,preprint,amsmath,amssymb,aps,prc]{revtex4-1}
\usepackage{float}
\usepackage{ulem}
\usepackage{graphicx}

\newcommand{\fref}[1]{Fig.\hspace{0.025in}\ref{#1}}

\newcommand{\eref}[1]{Eq.\hspace{0.025in}(\ref{#1})}

\begin{document}

\title{Noise and charge discreteness as ultimate limit for the THz operation of ultra-small electronic devices}

\author{Enrique Colom\'{e}s}
\affiliation{Departament d'Enginyeria Electr\`onica, Universitat Aut\`onoma de Barcelona, Spain}

\author{Javier Mateos}
\affiliation{Departamento de F\'{i}sica Aplicada, Universidad de Salamanca, Spain}

\author{Tom\'{a}s Gonz\'{a}lez}
\affiliation{Departamento de F\'{i}sica Aplicada, Universidad de Salamanca, Spain}

\author{Xavier Oriols}
\email{xavier.oriols@uab.cat}
\affiliation{Departament d'Enginyeria Electr\`onica, Universitat Aut\`onoma de Barcelona, Spain}




\begin{abstract}
To manufacture faster electron devices, the industry has entered into the nanoscale dimensions and Terahertz (THz) working frequencies. The discrete nature of the few electrons present simultaneously in the active region of ultra-small devices generate unavoidable fluctuations of the current at THz frequencies. The consequences of this noise remain unnoticed in the scientific community because its accurate understanding requires dealing with consecutive multi-time quantum measurements. Here, a modeling of the quantum measurement of the current at THz frequencies is introduced in terms of quantum (Bohmian) trajectories. With this new understanding, we develop an analytic model for THz noise as a function of the electron transit time and the sampling integration time, which finally determine the maximum device working frequency. The model is confirmed by either semi-classical or full- quantum time-dependent Monte Carlo simulations. All these results show that intrinsic THz noise increases unlimitedly when the volume of the active region decreases. All attempts to minimize the low signal-to-noise ratio of these ultra-small devices to get effective THz working frequencies are incompatible with the basic elements of the scaling strategy. One can develop THz electron devices, but they cannot have ultra-small dimensions. Or, one can fabricate ultra-small electron devices, but they cannot be used for THz working frequencies.
\end{abstract}

\maketitle

\thispagestyle{empty}


\section{Introduction}
\label{intro}

The main reasons for decreasing electron devices towards nanoscale dimensions are providing large scale transistor integration, lower power dissipation and high speed commutation \cite{moore}. Therefore, 3D structures like Fin-FETs or Gate-All-Around FETs based on Si nanowires (also on graphene or other 2D materials) are the typical ultra-small devices expected to play an important role in next-future electronics \cite{irds}. These ultra-small devices open new technological challenges that, step by step, are being properly solved (high-K dielectrics avoid spurious gate tunneling, multi-gate structures avoid short-channel effects, etc.). However, a new unexpected problem is presented in this paper for the operation of these ultra-small devices when approaching Terahertz (THz) working frequencies. The problem affects small-volume devices customarily developed by the electronics industry to continue with the ongoing scaling strategy, where the information is manipulated by means of electrical signals based on the motion of charge carriers and associated currents. All previous mentioned structures have an active region with a very small volume, with channel lengths and lateral dimensions of few nanometers. Thus, very few electrons are responsible for carrying the electrical current. We argue that the fluctuations of the current at THz frequencies make the predicted fast logic operation of these ultra-small devices inaccessible. The signal is defined as the part of the acquired current where  the information is encoded, while the noise is the difference between the current and the signal. We show that the THz noise grows when the volume of the active region decreases. Thus, if we keep a reasonable signal to avoid large power consumption, then the signal-to-noise ratio (SNR) at THz frequencies becomes intolerable for practical applications. Even avoiding all sources of noise that can be minimized by technological means, the noise that we are discussing in this paper will not diminish because it is just related to the discreteness of the electron charge. We emphasize that the relevance of our work resides on evidencing this noise limitation for nowadays technologies available in the electronic industry and providing the physical bases for the appropriate design of forthcoming generations of THz devices to elude this limit.

In spite of its obvious interest for the industry, very few papers analyze the behavior of the noise  of the electrical current in such ultra-small devices at THz frequencies. The reasons are the theoretical and computational difficulties that a proper study of THz noise in quantum devices has. Classically, the route to analyze THz fluctuations is unambiguously well-established, for example, through the successful Monte Carlo simulation of the Boltzmann equation for electrons. However, in principle, it is not obvious how semi-classical predictions can be extrapolated to the ultra-small devices mentioned above, where the wave nature of electrons becomes fundamental.  Most quantum electron device simulators are uniquely developed to study steady-state properties (the signal encoded in the DC),  which require much simpler theoretical and computational efforts than the study of the quantum fluctuations of the current (the noise). These conceptual and  computational difficulties explain why the THz noise limitation mentioned in this work has remained essentially ignored by the scientific community, in spite of its dramatic implications.

The THz noise restriction due to the discreteness of charge presented here has some similarities (and some differences) with the problem of the discrete doping. As it is well-recognized by the scientific community, when the number of dopants is very small, the intrinsic uncertainties in the fabrication process of the device implies important variations from one device to another. Therefore, the assumption of a continuous doping provides unrealistic predictions about the behavior of electron devices, because it ignores the large dispersion on the characteristics of the supposedly ``identical'' electron devices. Here, we show that when the number of electrons in the active device region is very small, then, the intrinsic uncertainties in the dynamics of electrons imply important variations in the electrical current at THz frequencies. Again, assuming a continuous flux of charge provides unrealistic predictions about the performance of these devices. Certainly, this noise disappears if the information about the signal is obtained after averaging the instantaneous current over times much larger than the typical electron transit time (as the problem of discrete dopants would easily disappear if an ensemble over different devices were allowed) at the price of renouncing to the expected speed of these ultra-small devices. Thus, the dramatic conclusions explained here are not relevant to the DC behavior of ultra-small devices. Our conclusions are only applicable to their high-frequency behavior. In other words, one can develop THz electron devices, but they cannot have ultra-small dimensions. Or, one can fabricate ultra-small electron devices, but they cannot work at THz frequencies.

\begin{figure}[H]
\centering
\includegraphics[width=0.6\columnwidth]{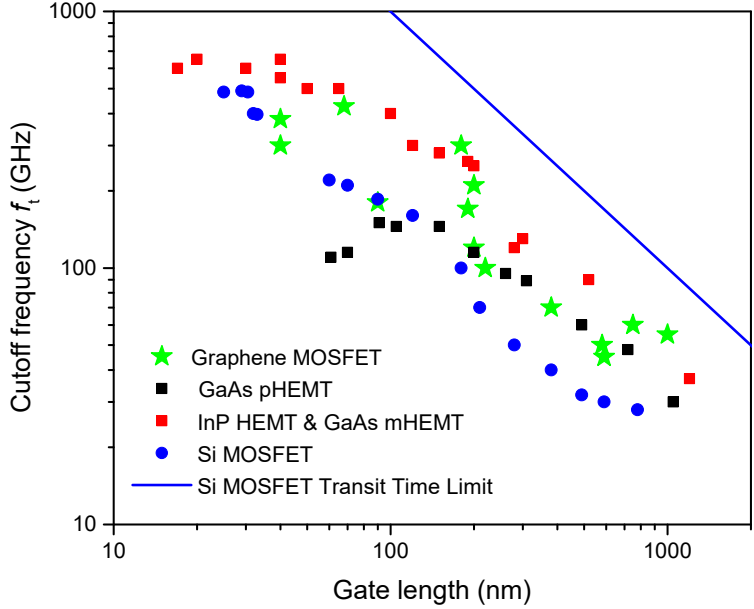}
\caption{Experimental cutoff frequency $f_t$ from a set of transistors based on different materials from Refs.\cite{figini1,figini2} as a function of their gate length. With blue solid line we plot the theoretical transit time limit (\eref{trlim}) for Si MOSFETs (velocity obtained from Ref.\cite{Rais}). We observe that $f_t$  is below the transit time limit $f_{\tau}$, specially for small gate lengths.}
\label{clock}
\end{figure}

During decades, the cutoff frequency of the transistors has been linked to their gate length by using the transit time limit \cite{trlim}. Thus, neglecting parasitic effects, the theoretical cutoff frequency $f_{\tau}$ is inversely proportional to the electron transit time $\tau$ and given by\cite{ft1}:
\begin{eqnarray}
f_{\tau} \leq \dfrac{1}{\tau}=\dfrac{v}{L}
\label{trlim}
\end{eqnarray}
with $\tau=L/v$ defined as the length $L$ of the active region in the transport direction divided by the average velocity $v$ of the electrons in this direction. In \fref{clock}, we plot the experimental cutoff frequency $f_t$  as a function of the gate length for several transistors based on different materials\cite{figini1,figini2}. The cutoff frequencies of all transistors follow the trend associated with the transit time limit. We plot as an example, the transit time limit for Si MOSFETs from \eref{trlim} with a solid blue curve (with $v$ obtained from Ref. \cite{Rais}), which is inversely proportional to the length of the active region. For small devices, \eref{trlim} provides unrealistic high cutoff frequencies. In this work, we discuss that apart from this transit time limit, there is another intrinsic limit, due to the discrete nature of electrons, that determines which is really the maximum working frequency of transistors. This discrete nature of electrons creates fluctuations in the current at times comparable to the transit time. We argue that these THz fluctuations will limit the device miniaturization for high frequency electronic applications. Smaller devices are certainly faster, but they are also nosier. Therefore, a trade-off between the desired speed and the acceptable noise is needed in ultra-small devices when increasing their working frequency.

\section{Total (displacement plus particle) current and noise in quantum electron devices}
\label{sec3}

The Monte Carlo technique applied to the solution of the Boltzmann equation has been the preferred tool to simulate electron devices during decades \cite{Montecarlo}. Through the explicit simulation of electron trajectories, it provides an intuitive and accurate simulation tool for predicting either static or dynamic properties of electron devices. In addition, because of its versatility, it has also been invoked as a ``simulated experiment" to save costs and efforts in the development of industrial prototypes of semi-classical electron devices. Because of miniaturization, the study of the dynamics of electrons inside ultra-small devices needs new concepts (like energy quantization and tunneling) linked to the wave nature of the electrons. For this reason, in the last fifteen years, a first revolution have taken place in the electron device modeling community moving from classical simulation tools to quantum ones (with more computational cost). Many different simulators has been successfully built during this time to compute the properties of ultra-small devices (NEMO \cite{sim6}, NEXTNANO \cite{sim8}, TiberCad \cite{sim9}, the NanoTCAD ViDES simulator \cite{sim9bis} or the Transiesta \cite{sim19}). These quantum simulators are basically devoted to static (DC) properties of nanodevices and therefore they are unable to properly predict the dynamics related to THz noise discussed in this work. As a byproduct of the present work, we also argue that a second revolution in the development of electron device simulators is needed to properly tackle the dynamic properties of this state-of-the-art ultra-small devices. There are two basic elements that justify the need for this second revolution and show its difficulties. 

First, the dynamic properties of electron devices are linked to time-correlations of the electrical current, which implies a proper modeling of the measurement process of the quantum device at different times. In a tunneling barrier, with equal transmission and reflection probabilities, we cannot say that half of the charge of a single electron is transmitted and half reflected. Each individual electron carries a charge equal to $q=-1.6\times10^{-19}$ C and it is either transmitted or reflected, but not both. The wave function solution of the Schr\"{o}dinger equation provides a \textit{natural} statistical view that explains that, for an ensemble average, half of the number of injected electrons are transmitted and half reflected, but such statistical view alone provided by the linear wave function is not enough to understand the partition noise created by the barrier on a single electron. A proper modeling of the collapse of the wave function, breaking the superposition of the wave function in left and right sides of the barrier, is needed to recover the discrete nature of charge of individual electrons at a quantum level.  In technical words, apart from the Schr\"{o}dinger equation, some type of modeling of the stochastic collapse law (reduction of state) in the quantum equation of motion of the electron is needed to go beyond DC predictions of quantum electron devices. 

Second, in fact, the discrete nature of electrons alone is not enough to understand the electrical current at THz frequencies. The relevant total current is the sum of the conduction (flux of particles) plus the displacement (time-derivative of the electric field) components \cite{ramo,shockley,ZhenTED,zhen2}. The displacement current on a surface is different from zero whenever electrons are able to modify the electric field on it (independently on how far the electrons are from the surface). Therefore, while in steady state (DC) conditions the displacement current is zero because of the time averaging, at high frequencies a proper self-consistent solution of Maxwell and transport equations is needed to know the interplay between scalar potentials and electron dynamics. In fact, under reasonable approximations, the electric field generated by electrons has to satisfy only the time-dependent Gauss law (with proper boundary conditions) plus the usual electron transport equation\cite{ft2}. In technical words, some type of modeling of the operator involved with the quantum measurement of the displacement current (not only with the quantum measurement of the particle current) is mandatory for THz predictions. 

The two above new ingredients required for the simulation of the electrical current at very high frequency seem to be not fully appreciated by the scientific community dealing with the simulation of ultra-small quantum devices. As we have commented, most quantum computational tools are devoted only to steady-state (DC) predictions, ignoring the displacement current and the multi-time measurement. In the literature, for general open quantum systems, there are basically two types of strategies to develop non-unitary equations of motion under multi-time (or continuous) measurement\cite{open}. The first strategy is developing equations of motion for the (reduced) density matrix. An example of this first type, valid for Markovian open systems only, is the Lindblad master equation\cite{lindblad}. The second strategy is to decompose (unravel) the density matrix in terms of individual (pure) states, and look for an equation of motion of such individual states. An example of this second type, valid for either Markovian or non-Markovian systems, is the stochastic Schr\"{o}dinger equations \cite{open,vega,SSE}. The main idea is finding the state solution of a Schr\"{o}dinger equation which includes the degree of freedom of the open system plus an external parameter representing the rest of degrees of freedom. Because of their dependence on such external parameter, these states are called conditional states (or conditional wave functions).  As explained recently\cite{Zhen, Devashish}, Gambetta and Wiseman\cite{Gambetta,wiseman} showed that the physical connection of a property of one conditional states between different times requires a quantum theory (like Bohmian mechanics) where the definition of a conditional state has a clear physical (not only mathematical) meaning. In this work, we will use the BITLLES simulator \cite{OriolsPRL,EnriquePRB,Oriols2013,BITLLES1,BITLLES2,BITLLES5,BITLLES6},  developed following this second strategy, to provide numerical support to the conclusions of THz noise in ultra-small electron devices.  The displacement current\cite{Zhen_thesis} and the back action induced by the continuous measurement of the electrical current\cite{DamianoPRL} are directly incorporated into the BITLLES simulator. In this work, we adapted  the previous BITLLES simulator to 2D linear band materials where the wave nature of electrons is described by a bispinor solution of the Dirac equation\cite{Enrique_thesis}. Next, before providing accurate numerical results of THz noise for graphene devices, we explain the main results of this work for very simplified electron device scenarios using trajectories. For those readers familiar with Monte Carlo simulations of the Boltzmann equation, the expressions developed here will seem quite trivial, but such expressions are also rigorously valid for the quantum regime, where such trajectories have to be understood as quantum Bohmian trajectories which, when properly including the measuring apparatus, exactly reproduce the quantum results.

The computation of the total current on a particular surface $S_i$ of the simulation box represented by \fref{fig1}, due to the time-dependent electric field generated by charge inside and outside of the active region and the (particle) classical or quantum current density due to electrons crossing the surface is:
\begin{equation}
I_i(t)=\int_{S_i} \vec J_c(\vec r,t) \cdot \mathrm d \vec s +\int_{S_i} \epsilon(\vec r)\frac{\mathrm d\vec E(\vec r,t)}{\mathrm d t}\cdot \mathrm d \vec s
\label{i1}
\end{equation}
where $\epsilon(\bar{r},t)$ is the (inhomogeneous) electric permittivity. The subindex $i$ indicates the surface $S_i$ where the current $I_i(t)$ is measured. Whenever not relevant in the discussion, the subindex $i$ and the time $t$ will not be indicated. The electrical field $\bar{E}(\vec{r},t)$ is solution of the Gauss equation to account for the Coulomb interaction among electrons, which is a huge computational problem in quantum systems (the many body problem \cite{libro}) requiring educated guesses. The current (particle) density $\bar{J}_c(\vec{r},t)$ is just a vector equal to the product of the electron charge density multiplied by the (classical or Bohmian) vector velocity of the electron. In the quantum case, this electron velocity includes all pure quantum (contextual, non-local) phenomena and the ensemble of $\vec J_c(\vec r,t)$ over many trajectories corresponds to the standard mean value of the quantum current operator \cite{libro}. 
 
Dealing with the instantaneous current $I_i(t)$ is just an idealization, and, in order to correctly reproduce the experimental conditions (in which an acquisition time is intrinsically involved), we compute a time-averaged value of the instantaneous current in the surface $S_i$ during the time interval $[t-T,t]$, defined as :
\begin{eqnarray}
I_{T,i}(t)=\frac{\int_{t-T}^t I_i(t')dt'}{T}
\label{i2}
\end{eqnarray}
where $T$ is the averaging time (equivalent to the acquisition or sampling integration time in a measurement) which limits the maximum working (or operating) frequency of the device. The standard deviation $\sigma_{T,i}$ of the averaged current $I_{T,i}(t)$ quantifies the noise of such a device:
\begin{eqnarray}
\sigma_{T,i}=\sqrt{\langle \triangle I_{T,i}^2\rangle}=\sqrt{var(I_{T,i})}
\label{nois}
\end{eqnarray}
Let us notice that the noise discussed in this work is completely suppressed when $T\rightarrow\infty$ in \eref{i2} because the current $I_{T\to\infty,i}=I_{DC}$ has no uncertainty. However, as we will demonstrate in this paper, increasing  $T$ drastically reduces the frequency of operation below the THz range. 
The SNR is the key parameter when characterizing the noise-related limit of operation of a given device, since it tells us how strong is the signal compared to the noise, and how much noise we can accept in our application. We can write the SNR, for each particular value of  $T$, as: 
\begin{eqnarray}
SNR_{T,i}=\frac{ I_{DC,i} }{\sigma_{T,i}}
\label{snr}
\end{eqnarray} 
where $ I_{DC,i} $ is the DC value of the current $I_{T\to\infty,i}=I_{DC,i}$ understood here as the signal (the part of $I_{T,i}$ that encodes the information, not the noise).

\subsection{The Ramo-Shockley-Pellegrini theorem}

\begin{figure}[H]
\centering
\includegraphics[width=0.6\columnwidth]{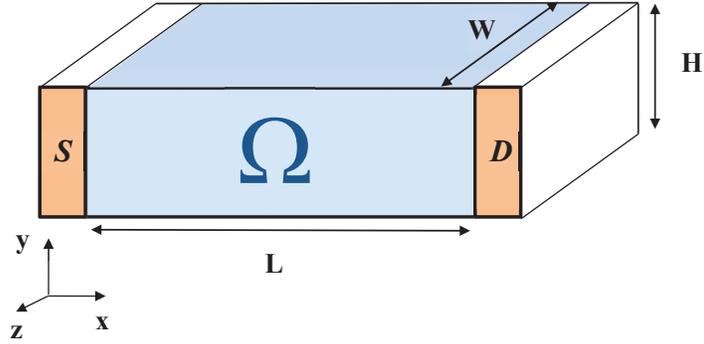}
\caption{Schematic representation of the simple idealized system used in this section to study the THz noise. The active region of the 2D FET is $\Omega=L \times H \times W$, being $L$ the length, $H$ the height of the channel and $W$ the width of the transistor.}
\label{fig1}
\end{figure}

In order to explain the importance of the discreteness of charge on the THz noise, in this section we study a very simple scenario: a two terminal device of length $L$ between two metallic contacts, represented by the source (S) and drain (D) contacts in \fref{fig1}. The volume of the active device region is $\Omega=L\cdot W\cdot H$. To simplify the discussion, electron transport is assumed to be fully ballistic in all simulations done in this work (which is a reasonable assumption for short-gate-length devices considered).

The Ramo-Shockley-Pellegrini theorem \cite{pellegrini} provides an alternative and useful expression for the total current appearing in \eref{i1}:
\begin{equation}
I_{i}(t)= -\int_\Omega \vec{F}_i(\vec{r}) \cdot \vec{J}_c(\vec{r}, t)  \mathrm{d}\nu + \int_{S_\Omega} \epsilon(\vec{r})  \frac{\mathrm{d}V(\vec{r},t)}{\mathrm{d}t}  \vec{F}_i(\vec{r}) \cdot  \mathrm{d}\vec{s}  
\label{i3}
\end{equation}
The surface $S_\Omega$ in \eref{i3} is now a closed surface enclosing an arbitrary volume $\Omega$ and $\vec{F}_i(\vec{r})$ is a mathematical vector field defined in Ref. \cite{zhen2}. It can be proven that in a two terminal device, the instantaneous time current assigned to a $k$-th electron while crossing the device with velocity $v_x^k(t)$ in the $x$ direction can be written as $I_i(t)=q\;v_x^k(t)/L$ (see Refs. \cite{libro} and  \cite {pellegrini}). Then, the time averaged current due to all electrons inside the device in \eref{i2} can be rewritten as:
\begin{equation}
I_{T}(t)=\frac {1} {T} \int_{t-T}^t dt \left(\dfrac{q}{L}\sum_{k=1}^{N_{\Omega}(t)}v^k_x(t)\right)=\dfrac{\dfrac{q}{L}\sum_{k=1}^{N_{\Omega}(t)} \Delta x^k_T(t)}{T} 
\label{i4}
\end{equation}
where $N_{\Omega}(t)$ is the number of electrons inside the volume $\Omega$ at time t. We have defined $\Delta x_T^k(t)=\int_{t-T}^t dt \;v^k_x(t)$ as the distance completed by this $k$-th electron during the time interval $t-T\leq t'\leq t$ inside the active region. Therefore, $0\leq\Delta x_T^k(t)\leq L$. 

It is important to remark that the contribution of an electron to the current in \eref{i4} is zero when the electron position is outside of the limits of the active region. This is because we assume that the density of electrons in the metallic contacts is so high that the electric field generated by one moving electron in the metal (outside the active region) is rapidly screened by the other (free) electrons in the metal without providing any displacement current. This is a fundamental element in our discussion, because it explains that the transfer of charge $q$ from left to right (or viceversa) can be understood as a current pulse during the transit time of the electron. The transmitted charge during this time is given by time-integrating \eref{i4} as $q=\int_0^{\tau} dt \;I_T(t)$. 

At this point two different scenarios can be distinguished, when $T$ is much shorter than the typical electron transit time $\tau$ (scenario ``a'') and when it is much larger (scenario ``b'').  In the next two subsections we develop \eref{i4} and its noise for these two different limits.

\subsection{Scenario a: $T$ much shorter than the transit time $\tau$ ($T \ll \tau$)}
\label{scea}

The first scenario corresponds to the case when the averaging time T is much shorter than the transit time $\tau$  of most of the electrons crossing the device, $T \ll \tau$. In that case, electrons are not able to cross the volume $\Omega$ during the time $T$.  For simplicity, in this preliminary analytic discussion, we assume a uniform velocity  $v_k^x (t)\approx v_e$ and $\Delta x_T^k(t)\approx \Delta x$ (in fact, this approximation is very accurate for linear band-structure materials such as graphene). Then, we have $\Delta x =v_e \cdot T$ and \eref{i4} can be rewritten as:
\begin{equation}
I_{T_a}=\dfrac{q}{L}v_e \langle N_{\Omega} \rangle
\label{i4bis}
\end{equation}
If we define $N_{cross,T}$ as the number of electrons crossing the whole device during the time interval $T$, in this scenario, we have $N_{cross,T}\ll N_{\Omega}$. To simplify the notation, whenever not relevant, we will omit the dependence on time of the parameters of the current. From \eref{nois}, the noise then is:
\begin{equation}
\sigma_{T_a}=\frac{q v_e}{L}\sqrt{var(N_{\Omega})}
\label{noisa}
\end{equation}

In order to understand better \eref{noisa}, let us take two different devices, the one we are interested in (with length $L$) and an arbitrary one (with length $L'$). Since electrons have no time to cross the device in both cases, different length devices imply a difference in the number of particles inside them. Because we consider a simplified scenario, where there is no correlation among electrons, $var(N_{\Omega})=\frac{L}{L'}var(N_{\Omega'})$. Then, we can rewrite \eref{noisa} as:
\begin{equation}
\sigma_{T_a} = \frac{q v_e}{L}\sqrt{\frac{L}{L'}var(N_{\Omega'})} = \frac{A}{\sqrt{L}}  
\label{noisa2}
\end{equation}
with $A$ being a constant (independent on $T$) which depends on $v_e$ and the topology of the devices. \eref{noisa2} indicates that when $T \ll \tau$, the noise is inversely proportional to the square root of the length of the device in the transport direction. A device with smaller $L$ provides more noise. The reason why an electron inside the active region (without reaching the contacts), provides current and charge fluctuations on the contact is because the original \eref{i4} includes the displacement current. Without the explicit consideration of such displacement current, this limit cannot be established.

\subsection{Scenario b: $T$ much larger than the transit time $\tau$ ($T \gg \tau$)}
\label{sceb}

When the averaging time is much larger than the transit time, electrons complete the distance $L$ during the time interval $T$, so $\Delta x_T=L$ in \eref{i4} and then the current is:
\begin{equation}
I_{T_b}=\dfrac{q}{TL} \langle N_{cross,T} \rangle L=\dfrac{q}{T} \langle N_{cross,T} \rangle 
\label{iTlimb}
\end{equation}
where we remind that $N_{cross,T}$  is the number of electrons crossing the device during the time interval $T$ (when $T \gg \tau$ the number of electrons crossing the device during the time interval $T$ is much larger than the instantaneous number of electrons inside, $N_{\Omega}\ll N_{cross,T}$). From \eref{iTlimb}, the noise is:
\begin{equation}
\sigma_{T_b}=\frac{q}{T}\sqrt{var(N_{cross,T})}
\label{noisb}
\end{equation}
Now, we will proceed similarly as before. But, let us remark that the situation now is different to the previous one. Then, we can establish a different time interval $T'$ (still $T' >>\tau$) so that $var(N_{cross,T})=\frac{T}{T'}{var(N_{cross,T'})}$, where $N_{cross,T'}$ is the number of electrons crossing the device during $T'$. Then, the noise of our device is:
\begin{equation}
\sigma_{T_b} = \frac{q}{T}\sqrt{\frac{T}{T'}var(N'_{cross,T'})}  = \frac{B}{\sqrt{T}}    
\label{noisb2}
\end{equation}
with $B$ being a constant independent on $L$, which again depends on the topology of the devices. From \eref{noisb2}, we see that effectively, in this limit, the noise is independent of the device length, but is inversely proportional to the square root of the averaging time $T$.

\subsection{Analytic maximum working frequency: the transit time limit or the noise limit?}

In this subsection, we show analytically that the maximum working frequency of state-of-the-art ultra-small devices is not always determined by the transit time limit, but by the new noise limit discussed here.  Clearly, for digital electronics, the limit imposed by the transit time $T \ll \tau$ (scenario a) cannot be overcome, i.e. a device cannot work at frequencies higher than the ones imposed by the transit time in \eref{trlim}. However, we argue in this paper that the maximum working frequency of many nanoscale devices is, in fact, determined by the noise limit, not by the transit time limit.

Let us derive analytically what is the noise limit imposed for $T \gg \tau$ (scenario b)  given by \eref{noisb} for a 2D and a 3D device. For that purpose, using \eref{iTlimb} and \eref{noisb}, the SNR in \eref{snr} can be rewritten as:
\begin{equation}
SNR_{T_b}=\frac{  I_{DC,i}  }{\sigma_{T}}=\frac{\langle  N_{cross,T} \rangle}{\sqrt{var(N_{cross,T})}}
\label{snr2bis}
\end{equation}
There is a strong link between experimental averaging time $T$ and the amount of noise in \eref{snr2bis}. If we fix the amount of acceptable noise for a given circuit application, then $T$ must be increased up to reach the desired value of SNR. We define $T_{SNR_0}$ as the averaging time that satisfies the required signal-to-noise ratio value $SNR_0$. Therefore, we can define the noise-related working frequency limit as: 
\begin{equation}
f_n=1/T_{SNR_0}
\label{tsnrlim}
\end{equation}
We argue that, in many scenarios involving ultra-small devices, the noise limit in \eref{tsnrlim} gives a lower maximum working frequency than the transit time limit in \eref{trlim}.

To provide a compact expression relating $f_n$ and $f_{\tau}$, let us define $N_{cross,\tau}$ as the number of electrons crossing the device in the time interval $\tau$, then, using \eref{iTlimb} and \eref{noisb2} for the fixed value $SNR_0$, we get\cite{ft3}:
\begin{equation}
SNR_{0}=\frac{ I_{DC,i} }{\sigma_{T_b}}=\left({\frac{q\langle N_{cross,\tau} \rangle}{\tau}}\right)\bigg/\left({ \frac{q}{\sqrt{\tau}}\sqrt{var(N_{cross,\tau})}\frac{1}{\sqrt{T_{SNR_0}}}}\right)=\frac{\sqrt{T_{SNR_0}} }{\sqrt{\tau}}\frac{ \langle N_{cross,\tau} \rangle}{\sqrt{var(N_{cross,\tau})}}
\label{snr2}
\end{equation}

We can now obtain the ratio between the noise frequency limit  $f_n$ and the transit time frequency limit $f_{\tau}$ to verify which one is more relevant in determining the maximum working frequency of ultra-small devices.  We assume a Poisson probability distribution for carrier injection with a probability of success  $p$. We consider that $N_{inj,\tau}$ electrons attempt to be injected during the time $\tau$, then $\langle N_{cross,\tau} \rangle=var(N_{cross,\tau})=pN_{inj,\tau}$. From \eref{snr2} and \eref{trlim}, with $f_n=1/T_{SNR_0}$, we can straightforwardly obtain the ratio between $f_n$ and $f_{\tau}$  as:
\begin{equation}
\frac{f_n}{f_{\tau}}=\frac{1}{SNR_0^2}pN_{inj,\tau}
\label{rf}
\end{equation}

In the case of a 3D device, the number of electrons attempting to be injected (with electrons going just in one direction) from the phase-space density is $N_{inj,\tau}$ $=L W H\frac{k_f^3}{6\pi^2}$  with $k_f$ the Fermi wave vector (we have already taken into account the spin degeneracy). Then, from \eref{rf} :
\begin{eqnarray}
\frac{f_n}{f_{\tau}} = \dfrac{L W H k_f^3}{6\pi^2 SNR_0^2}p
\label{nlim3d}
\end{eqnarray}

In the case of a device whose channel is a 2D material (such as graphene), the number of electrons  is $N_{inj,\tau}=L W\frac{k_f^2}{4\pi} $. Then, again, from \eref{rf} we obtain:
\begin{eqnarray}
\frac{f_n}{f_{\tau}}= \dfrac{L W k_f^2}{4\pi SNR_0^2}p
\label{nlim2d}
\end{eqnarray}
We remark that whether the condition $f_n/f_{\tau}< 1$ is fullfilled (meaning that the $f_n$  limit is reached at a lower frequency than the limit due to $f_{\tau}$), or the opposite one $f_n/f_{\tau}> 1$, depends strongly on the transistor characteristics. In any case, as a general trend, we see from \eref{rf} that the lower amount of electrons present in the active region, the lower the value of the ratio $f_n/f_{\tau}$. Therefore, the noise limit discussed here is more and more relevant as the dimensions of electron devices become smaller and smaller. When using the planar 2D MOSFET architectures, the value of the current could be increased (and therefore the $f_n/f_{\tau}$ factor) by increasing the device width $W$, but with present-day technologies this solution is much more complex (i. e. parallel fins or nanowires have to be added in FIN-FETs or GAA-FETs, respectively). 

\begin{figure}[H]
\centering
\includegraphics[width=0.7\columnwidth]{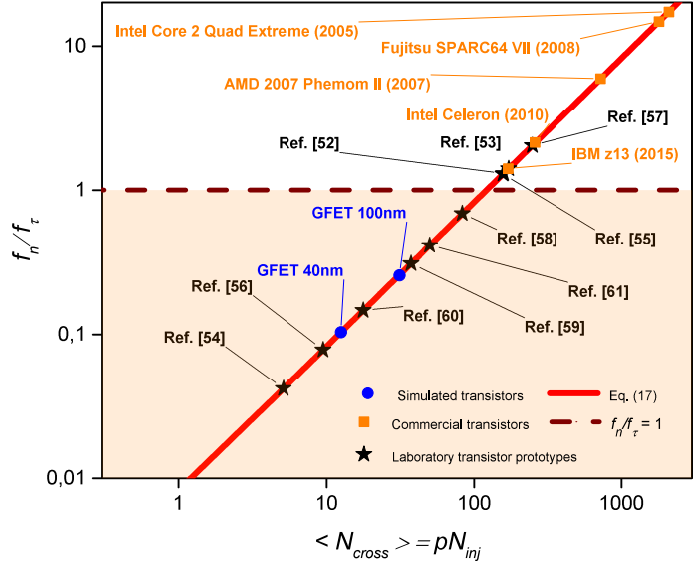}
\caption{The ratio ${f_n}/{f_{\tau}}$ is plotted as a function of the number of transport electrons inside a transistor. With a red solid line we plot \eref{rf} and with a brown dashed line when the ratio ${f_n}/{f_{\tau}}$ is equal to one.  Points corresponding to commercial transistors, laboratory transistor prototypes and the devices that will be simulated in next section (whose details are given in table \ref{table1})  are plotted with symbols. We see how  the ${f_n}/{f_{\tau}}$ ratio is lower than one for many of these transistors, indicating that the noise limit is relevant nowadays.}
\label{fig_ana}
\end{figure}

We wanted to test if existing transistors have already entered into the regime where the maximum working frequency is limited by the noise, and not by the transit time. For that purpose, in \fref{fig_ana}, the ratio of ${f_n}/{f_{\tau}}$ is shown for different scenarios and transistors. We represent with a red solid line \eref{rf}, and with a brown dashed line when the ratio is equal to one. For values lower than one (shaded region) the working frequency is limited by the noise, otherwise it is limited by the transit time limit. Different transistors are plotted: some laboratory prototypes (black star symbols), some commercial transistors, which already appeared in \fref{clock} (orange square symbols), and two GFET transistors, that will be simulated in next section (blue rounded symbols). All transistor ratios where obtained through \eref{nlim3d} and \eref{nlim2d}. In these expressions, we estimated $p=0.3$ from the comparison of analytic and computed results of the GFET transistors simulated in next section. We accept as tolerable noise a SNR equal to 11. (see Ref. \cite{lazslo0}). This is the minimum SNR (and associated maximum noise level) that can be accepted in a logical device for tolerable errors. 

\begin{table}

\begin{tabular}[b]{| c | c | c | c | c | c | c | c | }
\hline
 & Type & $W$ $(nm)$ & $L$ $(nm)$ & $H$ $(nm)$ & $k_f$ $(nm^{-1})$  & Dimensionality & ${f_n}/{f_{\tau}}$ \\
\hline
Ref. \cite{fefet} & FeFET (Si) & 80 & 20 & 7 & 1,41 &\eref{nlim3d} (3D)&  1.30 \\
\hline
Ref. \cite{cn} & CNT & 9,42 & 32 & - & 4,88 &\eref{nlim2d} (2D)& 1.41 \\
\hline
Ref. \cite{cn2} & CNT & 4,08 & 10 & - & 2,29  &\eref{nlim2d} (2D)& 0.04\\
\hline
Ref. \cite{cn3} & CNT & 62,83 & 20 & - & 2,29 &\eref{nlim2d} (2D)& 1.29\\
\hline
Ref. \cite{nw1} & Nanotube(Si) & 10 & 20 & - & 1,41 &\eref{nlim2d} (2D)& 0.07\\
\hline
Ref. \cite{nw2} & Nanotube (Si) & 25,13 & 150 & - & 1,66 &\eref{nlim2d} (2D)& 2.04 \\
\hline
Ref. \cite{nw3} & Nanowire (Ge/Si) & 44 & 40 & - & 1,41 &\eref{nlim2d} (2D)& 0.68\\
\hline
Ref. \cite{gaat} & GAA FET & 44 & 12 & 5 & 1,41 &\eref{nlim3d} (3D)& 0.30\\
\hline
Ref. \cite{mos2} & MOS2 & 50 & 7,5 & - & 1,41  &\eref{nlim2d} (2D) & 0.14\\
\hline
Ref. \cite{grt} & GFET & 1000 & 40 & - & 0,22  &\eref{nlim2d} (2D)& 0.41 \\
\hline
GFET 40 nm & Simulated & 250 & 40 & -  & 0,22 &\eref{nlim2d} (2D)& 0.10\\
\hline
GFET 100 nm & Simulated & 250 & 100 &  -  & 0,22  &\eref{nlim2d} (2D) & 0.25\\
\hline
IBM z13 2015 & Commercial & 22 & 25 & 22 & 1,41   &\eref{nlim3d} (3D) & 1.41\\
\hline
AMD  2007 Phenom II & Commercial & 45 & 25 & 45 & 1,41 &\eref{nlim3d} (3D) & 5.90 \\
\hline
Fujitsu SPARC64 VII & Commercial & 65 & 30 & 65 & 1,41  &\eref{nlim3d} (3D) & 14.79\\
\hline
Intel Celeron & Commercial & 32 & 18 & 32 & 1,41 &\eref{nlim3d} (3D) & 2.15 \\
\hline
Intel Core 2 Quad Ext & Commercial & 65 & 35 & 65 & 1,41  &\eref{nlim3d} (3D) & 17.26 \\
\hline
\end{tabular}
\label{table1} 
\caption{ Table with the data of $W$, $L$, $H$ and Fermi wave vector $k_f$ indicating the use of \eref{nlim3d} (3D) or \eref{nlim2d} (2D) to evaluate the value of the ratio ${f_n}/{f_{\tau}}$ plotted in figure \ref{fig_ana} for commercial transistors (data obtained from \textit{http://cpudb.stanford.edu}), laboratory transistor prototypes (data obtained form the references) and the devices that will be simulated in next section. We estimated $p=0.3$ and SNR equal to 11. (see Ref. \cite{lazslo0})}

\end{table}

We observe in \fref{fig_ana} that many transistors are located in the shaded region ${f_n}/{f_{\tau}} < 1$ where the working frequency is limited by the noise limit, and not by the transit time limit. It is important to notice that, until now, the noise limit was not a problem. Nowadays, there is also a frequency limit imposed by dissipation that is well below the transit time and the noise limits discussed here. The power dissipation is directly proportional to the working frequency, i.e., the higher frequency we want to work, the more dissipation will occur. Thus, the overall amount of power that can be dissipated from the chip imposes a limit in the operating frequency on each transistor. Its is expected that this dissipation limit will be overcome with new strategies and technologies \cite{pl1,pl2}. Then, the transit time and noise limit will determine the intrinsic working frequency limit of ultra-small devices. Most commercial transistors have ratios ${f_n}/{f_{\tau}} > 1 $ and they are still not limited by the noise, but by the transit time. However, since transistor sizes are decreasing, less and less electrons are present in the device, and the noise limit becomes more and more relevant.  

\section{Numerical simulations for a simple two-terminal device}
\label{sec4}

In this section we present different numerical results corroborating the previous analytic predictions. Let also remark that all the expression presented previously are independent if we are in a classical or quantum regime. For semi-classical modeling, the electron trajectories appearing in \eref{i4} are computed from the semiclassical Monte Carlo solution of the Boltzmann equation, while for quantum modeling, the quantum trajectories are computed from a quantum time-dependent Monte Carlo BITLLES simulator where the electron velocity is computed from the (conditional) bispinor solution of the Dirac equation, which includes all quantum (non-classical) phenomena. One of the big merits of this work is to tackle the classical and quantum problem of the THz noise in ultra-small devices with the same language: electron trajectories. This fact greatly contributes to an easy and rigorous understanding of the problem and of its practical consequences for the future of ultra-small electron devices at THz frequencies. In the next two subsections we present semiclassical and quantum numerical results.

\subsection{Semiclassical numerical simulations}

Firstly, we present semiclassical Monte Carlo simulations for a two terminal device \cite{tj1,tj2,tj3}.  Inside the device, transport is assumed to be ballistic without electron-phonon collisions. A parabolic energy-band with an effective mass $m^*=0.25m_0$, being $m_0$ the free electron mass, is considered. For all simulations, a lattice temperature $T_{lat}=300$ K is considered. The variations in the number of particles inside the device come from the randomness of energies and times of entrance of electrons injected from the contact into the active region, following Fermi-Dirac statistics. In the literature, the fluctuations due to this randomness are known as thermal noise \cite{tj1}. The average charge density of carriers in the contacts is given by $n=10^{15}$ m$^{-3}$. The simulation time step is $dt=5\times10^{-15}$ s and the spatial grid is $dx=20$ nm. A self-consistent solution of the electric field and electron charge is established through the numerical solution of the Gauss (first Maxwell) law.  

\begin{figure}[H]
\centering
\includegraphics[width=0.7\columnwidth]{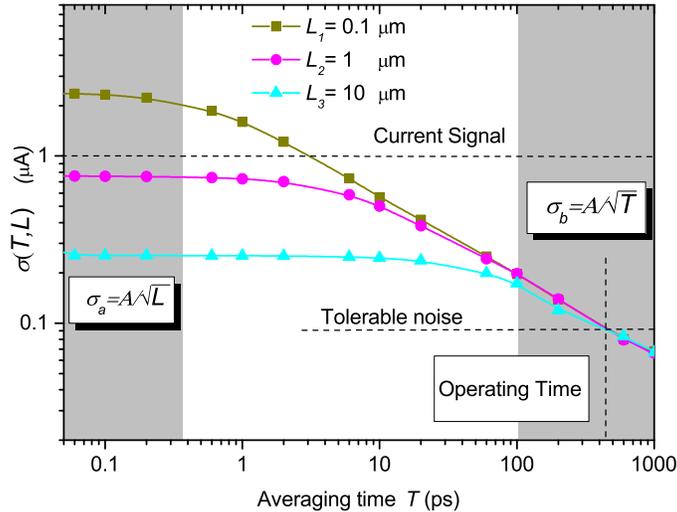}
\caption{Noise as a function of the averaging time $T$ for a two-terminal device with three different lengths $L$ when no drain-source bias is applied, $V_{DS}=0$ Volts.  Simulations were performed with the semiclassical Monte Carlo approach described in the text. In all three simulated devices, we consider $W \times H= 0.25\times 10^{-5}$ m$^2$. We accept as tolerable noise a $SNR$ equal to 11. See Ref.$\;$\cite{lazslo0}. }
\label{fig61}
\end{figure}

In \fref{fig61}, the value of $\sigma_{T}$ computed from Monte Carlo simulations using \eref{nois} is plotted. Three different device lengths  $L$ are studied.  For simplicity, injection from one of the contacts is just considered without bias applied (these simplifications will be avoided in next figure). The limits $\sigma_{T_a}$ and $\sigma_{T_b}$ are clearly reproduced in \fref{fig61}. Notice the dependence on $1/\sqrt{L}$ for $T<<\tau$ and the dependence on $1/\sqrt{T}$  for  $T \gg \tau$, as indicated in \eref{noisa2} and \eref{noisb2}, respectively. 

Let us now imagine that we design a device with $L=100$ nm for very high-frequency applications with an expected average-time interval of $T=\tau=1$ ps (i.e., an operating frequency of $1$ THz). Imagine that the design has a signal current value of $\langle I \rangle_{DC}=1$  $\mu$A (horizontal dashed line in \fref{fig61}) and that our particular application requires a typical factor 11 for the SNR (see tolerable noise in the horizontal dashed line in \fref{fig61}). Thus, we conclude, that the expected length $L=100$ nm  and operating time $T=1$ ps are incompatible with the required level of noise $\sigma_{T}=0.09$ $\mu$A. Such noise level can only be obtained working at $T=500$ ps (see vertical line in \fref{fig61}) where the three different lengths provide the same noise level. In conclusion, at the end of the day, there is no reason to prefer the shorter device. The larger one is equally valid. Let us remind that increasing the value of the current signal is not a generally acceptable solution because low power consumption is also a mandatory requirement to avoid dissipation in ultra-small devices.

\begin{figure}[H]
\centering
\includegraphics[width=0.7\columnwidth]{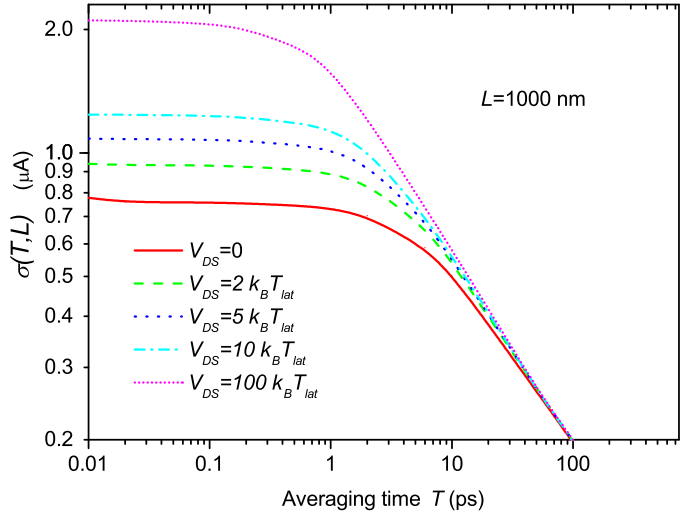}
\caption{Noise as a function of the averaging time $T$ for the device  of $L=1000$ nm and parameters used in the simulation of \fref{fig61}, when different applied drain-source voltages $V_{DS}$ are considered (whose values are written in terms of the Boltzmann constant $k_B$ and the lattice temperature $T_{lat}$). Simulations were performed with the semiclassical Monte Carlo approach described in the text. In all three simulated devices, we consider $W \times H= 0.25\times 10^{-5}$ m$^2$.}
\label{fig62}
\end{figure}

The ratio of two noise values corresponding to different lengths can be done with \eref{noisa2} and is equal to ${\sigma_{1}(t)}/{\sigma_{2}(t)}=\sqrt{{L_2}/{L_1}}$, which can be compared with the numerical data presented in \fref{fig61}. So, for $L_1=0.1$ $\mu$m and $L_2=1$ $\mu$m the ratio should be $\sqrt{{L_2}/{L_1}}=3.16$. According to the numerical results, the ratio is $3.09$. Therefore, analytic and numerical results fit quite good, showing the accuracy of the analytic results presented. The same calculus can be done with the other device ($L_3=10$ $\mu$m), showing the same accuracy.

In \fref{fig62}, we plot the same information as in \fref{fig61} for several applied drain-source bias $V_{DS}$. The consideration of far from equilibrium conditions does not change the previous overall conclusion (the bias conditions only modifies the quantitative values). We notice in \fref{fig62} that, for small averaging times $T<<\tau$, the value of $\sigma_{T_a}$ grows when larger bias is considered because the (mean) velocity of electrons, $v_e$, present in \eref{noisa2}, increases with bias. For the same reason, the regime $T>>\tau$ is reached for shorter $T$ as the bias is increased.

\subsection{Quantum numerical simulations}

Now, we will present similar numerical results as the ones showed previously, but we will simulate a graphene electron devices  (instead of a Silicon ones) with the use of the quantum simulator BITLLES\cite{OriolsPRL,EnriquePRB,Oriols2010,Oriols2013,BITLLES1,BITLLES2,BITLLES3,BITLLES5,BITLLES6}, which uses the quantum (Bohmian) trajectories applied to time-dependent  electron quantum transport\cite{ft4}. 

As it is well known graphene is a 2D material that because of its fascinating large electron velocity, many efforts have been done to study its real application in practical circuits. Since the beginning, many relevant voices in the literature \cite{figini1} have questioned the real potential utility of graphene as a useful semiconductor for logic gates since it is a gapless semiconductor (with many difficulties to provide low enough OFF currents). The literature is also studying other graphene structures like bilayer graphene (two coupled single graphene layers stacked as in graphite) \cite{bilayer} or strained graphene (mechanical deformation of the atomic structure) \cite{strain} to provide an energy gap between the conduction and valence bands. The ability of getting a gap in graphene by different means comes at the price of reducing its original extraordinarily high mobility. Other 2D materials are also intensively studied as potential candidates for future electronic technologies. In any case, our aim in the paper is not to construct a commercial transistor with a single graphene sheet, but to prove that for state-of-research devices fully based on quantum phenomena (like the single layer graphene devices simulated here based on Klein tunneling or any other prototype built from 2D structure) this noise limit will exist also, as well as it exists in more standard semiconductors.

\begin{figure}[H]
\centering
\includegraphics[width=0.7\columnwidth]{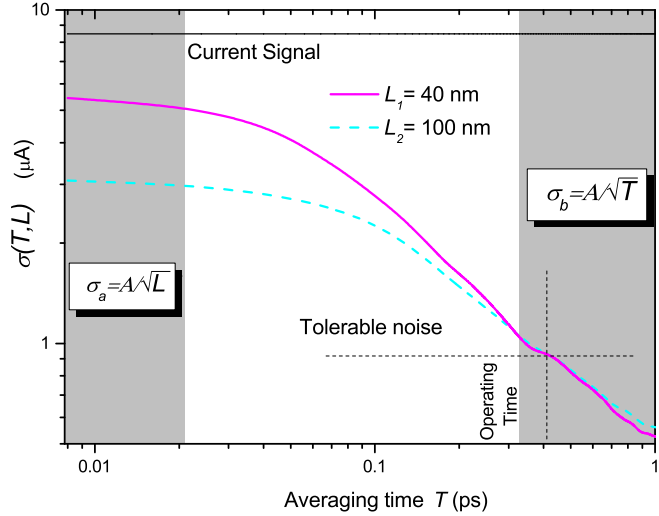}
\caption{Noise as a function of the averaging time $T$ for two-terminal graphene devices of different length when there is no  drain-source bias is applied, $V_{DS}=0$ Volts. Simulations were performed with the fully quantum BITLLES simulator. We accept as tolerable noise a SNR equal to 11 (see Ref.\cite{lazslo0}). The width of both transistors is $W=250$ nm and the Fermi level is set at $E_f=0.15$ eV.}
\label{fig62q}
\end{figure}

As explained in the Appendix, the quantum dynamics of an electron in graphene is given by the Dirac equation. The (conditional) wave function associated to each electron  is no longer a scalar, but a bispinor. Each electron is associated to a Bohmian trajectory computed from the wave function solution of the time-dependent Dirac equation. The initial state of each electron is a bispinor Gaussian wave packet defined deep inside the contacts with well-defined mean momentum (see the Appendix for more details). This quantum-trajectory formalism can be considered as the \textit{natural} quantum extension of the semiclassical Monte Carlo method mentioned before for classical systems. Electrons are injected following a binomial distribution according to the Fermi statistics. It includes the Coulomb interaction through the time-dependent solution of the Poisson equation with Dirichlet boundary conditions in the metals (contacts, gates) and Neumann ones in the rest of the surfaces. In the simulations, the spatial grid was set to $dx=dz=1$ nm and the time step $dt=10^{-16}$ s.  As argumented for the semi-classical simulations, ballistic transport is assumed as it is the expected transport regime in ultra-small graphene devices and a lattice temperature $T_{lat}=300$ K is considered.  Linear band structure, with constant velocity independently of the electron energy given by the Fermi velocity $v_f=10^6$ m/s is considered. First, we performed quantum simulations similar to the semi-classical ones presented in \fref{fig61}, injecting just from one side without applying any bias. 

In \fref{fig62q}, differences appear regarding the values of the current and noise with respect to the results in \fref{fig61}. This is because graphene is a linear band structure material, and therefore it has a constant velocity (independently of the electrons energy) whose value $v_f=10^6$ m/s is high compared to the typical ones in Silicon. Since the current is proportional to the carrier velocity, as it can be seen from \eref{i4}, current values in graphene are higher than the usual ones in the typical semiconductor devices. In addition, we see that the averaging times $T$ are much shorter, since the transit times $\tau$ are also much shorter (the devices are smaller and the carriers velocity is higher). This fact makes that the operating time for this graphene two-terminal idealized device is much smaller, $T \approx 0.5$ ps, with a theoretical cutoff frequency around $f \approx 2$ THz for this simplified scenario.

Apart from this difference, the shapes of \fref{fig61} and \fref{fig62q} are very similar. We see that in both of them, for averaging times smaller than the transit time, noise scales as $\sigma_{T_a} = {A}/{\sqrt{L}}$, whereas for averaging times larger than the transit time, $\sigma_{T_b} = {B}/{\sqrt{T}}$. Therefore, even when accounting for quantum effects, it can be seen that \eref{noisa2} and \eref{noisb2} are completely valid too. We compute the ratio between the noise for different device lengths when the $T$ is much shorter than the transit time. Similarly as done before, the ratio of the noise of the devices is $\sqrt{{L_2}/{L_1}}=\sqrt{{100}/{40}}=1.58$ and regarding the simulations, this ratio is $1.74$, showing again a reasonable agreement.

\section{Numerical simulation for a three terminal quantum GFET transistor}
\label{sec5}

In this section, we test the previous predictions about the THz noise for a realistic ultra-small device, without most of the simplifying assumptions that we have used in the analytical and previous simulations sections. We consider a graphene double gate transistor (source, drain, bottom and top gates) as the one depicted in \fref{fig1bis}. As explained in the Appendix and in the previous subsection, again, the quantum dynamics of electron in graphene is given by a Bohmian trajectory associated to the bispinor solution of the time-dependent Dirac equation. The injection of electrons is performed from both sides, source and drain, according to the (quasi) Fermi-Dirac statistics. There are two gates that affect the electric field inside the active device region, which have a strong influence on the transport along the channel and on the total current conservation. Now, the expression in \eref{i4} is no longer valid because it was developed for two-terminal devices. New volume $\Omega$ and function $\vec{F}_i(\vec{r})$ are needed in \eref{i3}. In any case, a peak of displacement current appears in the drain every time that an electron (Bohmian trajectory) traverses the channel. A similar peak appears in the displacement current of the source with some delay.  Finally, the displacement current in the gates will be of such shape that the sum of the total currents in the three terminals is zero at every time \cite{ZhenTED}. The relevant physics for the THz noise computation does not change significantly, and we expect the same qualitative results here.

\begin{figure}[H]
\centering
\includegraphics[width=0.5\columnwidth]{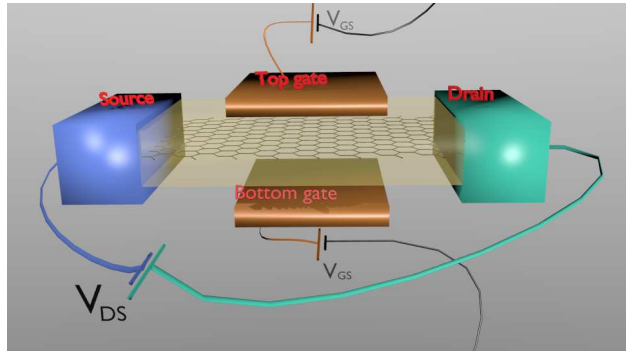}
\caption{Schematic representation of a double-gate graphene transistor in the BITLLES simulator. The channel (in this case graphene) is sandwiched between two dielectrics. The active region of the dual-gate 2D Fet is $\Omega=L \times (H'+H+H') \times W$, being $L$ the gate length, $H'$ the height of the dielectrics, $H$ the height of the channel and $W$ and the width of the transistor.}
\label{fig1bis}
\end{figure}

\begin{figure}[H]
\centering
\includegraphics[width=0.6\columnwidth]{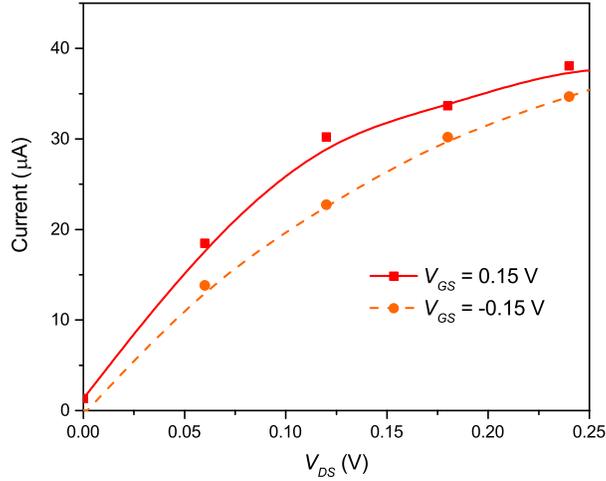}
\caption{Current-voltage characteristics for the a double-gate graphene transistor  whose active region volume is $\Omega=20 \times (5+1+5) \times 250$ nm$^3$. The optimum values for maximizing the different current levels are  $V_{DS}=0.12$ V and $V_{GS}=\pm 0.15$ V. The Fermi energy is $E_f=0.15$ eV.}
\label{vds6}
\end{figure}

In particular, we study how the THz noise affects our ability to distinguish between two current levels. For that purpose, we will establish a constant bias between the drain and source and we will change the gate voltage bias at some particular time. Then, we will obtain the total (particle plus displacement) current (obtained with the most general expression of the Ramo-Shockley-Pellegrini theorem, \eref{i3}) through \eref{i2} and estimate the minimum averaging/acquisition time that allows us to establish a difference between both states.

In order to establish the best value of the drain-source bias ($V_{DS}$) and the top and bottom gate values ($V_t=V_b \equiv V_{GS}$) to perform the transient, we made different current-voltage characteristic curves for different gate values. Among all curves, we chose the values that maximizes the differences between the drain-source currents that we will consider as the logical information $'1'$ that we refer here as level 1 (L1) state and the one that we consider the logical information of $'0'$ that we refer here as level 2 (L2) state. The transistor in \fref{fig1bis} has a volume $\Omega=20 \times (5+1+5) \times 250$ nm$^3$ and a device length in the transport direction $L=L_x'+L_x+L_x'= 40$ nm. Results are plotted in \fref{vds6}. There we see that the maximum difference between currents is achieved for a value of $V_{DS}=0.12$ V and $V_{GS}=\pm 0.15$ V. Therefore, L1 corresponds to $V_{GS}=0.15$ V, while L2 corresponds to  $V_{GS}=-0.15$ V. As it is well known, because of the presence of Klein tunneling, the graphene transistor cannot be switched off by any gate bias.

\begin{figure}[H]
\centering
\includegraphics[width=0.7\columnwidth]{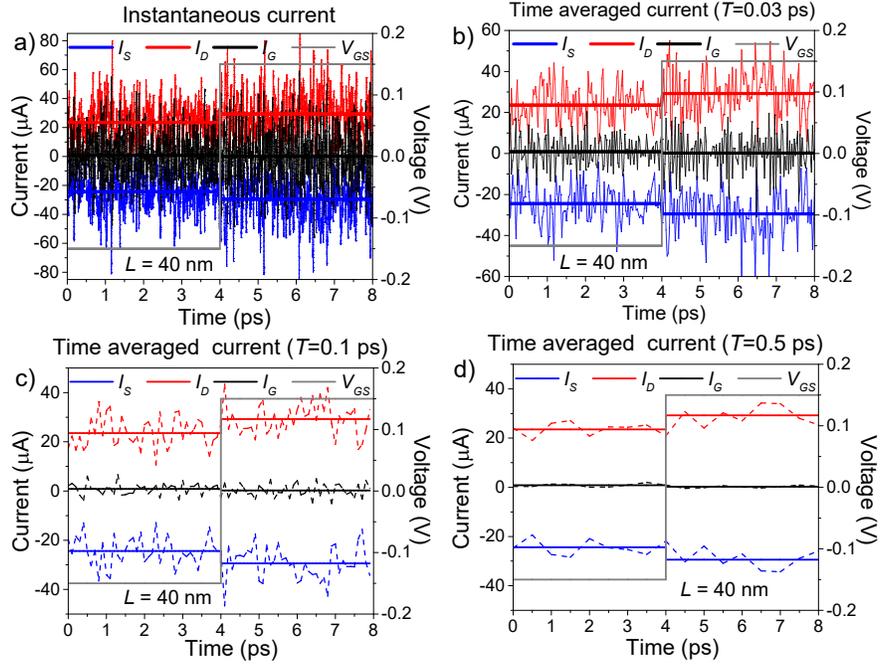}
\caption{Time-dependent currents of the a double-gate graphene transistor. a) Instantaneous current (time-averaged at the simulation step $dt=10^{-16}$ s) and its mean value as a function of time. Fluctuations do not allow to distinguish between L1 and L2. The  gray  line shows the gate voltage as a function of time. b) Averaged current through an averaging time $T=0.03$ ps. Still, we cannot distinguish between both levels. c) Averaged current through an averaging time $T=0.1$ ps. Noise decreases, but still  too high to distinguish both levels. d) Averaged current through an averaging time $T=0.5$ ps. Now, we can distinguish between both levels. This averaging time corresponds to the operating time shown in \fref{noistom6}.}
\label{inst6}
\end{figure}

Next, we made a simulation and analyze the switching times obtained. Initially, the gates value is $V_{GS}=-0.15$ V. After $4$ ps, the gate value is changed to $V_{GS}=0.15$, as shown in \fref{inst6}. The instantaneous current and gate voltage is plotted as a function of time in \fref{inst6}a). Current increases when switching the gate voltage. Clearly, without time averaging the current (with the use of \eref{i2}), noise does not allow us to differentiate L1 and L2. The question now is from which $T$ we can affirm that we are able to distinguish both states. In \fref{inst6}d), we present the averaged current for the same device with an averaging time $T=0.5$ ps. Now, noise allows us to distinguish between both states. We remark that the results presented here will not be obtained from an ensemble average over different experiments, since when an electron device is working in a real application, there is no interest in mean values of the current in different experiments, we are just interested on the time interval that the measurement equipment needs to clearly discern if our single electron device is in L1 or in L2. In more technical words, no ergodic argument can be invoked in the type of THz scenarios described here.  

\begin{figure}[H]
\centering
\includegraphics[width=0.6\columnwidth]{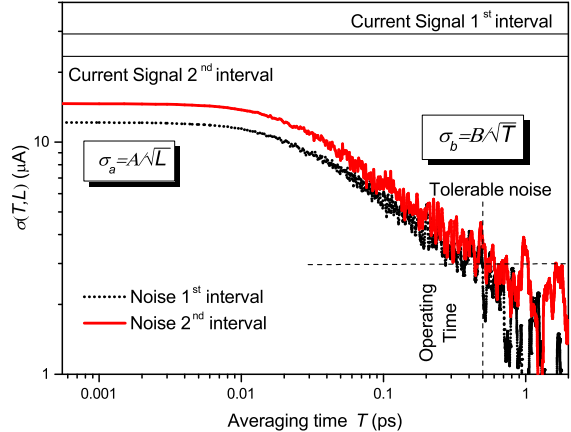}
\caption{Noise of the a double-gate graphene transistor as a function of the averaging time $T$ for the two time intervals of \fref{inst6}. We accept as tolerable noise a SNR equal to 11. See Ref.  \cite{lazslo0}. The applied source-drain voltage is $V_{DS}=0.12$ V.}
\label{noistom6}
\end{figure}

With this information, we can obtain the time averaged current and its associated noise (in the same way as it was done in \fref{fig62q}) for both time intervals (before and after switching the gate voltage). Results are plotted in \fref{noistom6}. Differently from \fref{fig62q}, results are noisier for large averaging times. This is because in \fref{fig62q} we averaged the results through different simulations in order to see very clearly the noise values. In this case, since we are interested just in what occurs in one experiment, we did not make the averaging between different simulations and results are noisier. Even if this case is very different from the one studied in \fref{fig62q} (there was no applied bias and there were no gates in the previous studies), we can still recognize the two different scenarios, $T << \tau$ and $T >> \tau$, and only the particular scenario depicted in \fref{inst6}(d) is acceptable. Therefore, as expected, all our previous predictions are still present in this realistic device. In summary, once we fix the amount of tolerable THz noise that our device application can accept, the lowest acceptable acquisition time (or the highest acceptable working frequency) is determined. Making measurements with a lower acquisition/integration time (or with a higher working frequency) to get a faster application would imply an intolerable THz noise.

\section{Conclusions}
\label{sec6}

One of the main interests for minimizing the size of electron devices is to perform applications working at higher and higher frequencies, until reaching THz working frequencies. Smaller transistors, in principle, imply to be able of working at higher frequencies since electrons need less time to travel through the device. How small can the active device region become? How high can be the associated working frequency? In this work, we see that, due to discreteness of charge, there is a new fundamental (noise) limit (apart from the transit time limit) for the strategy of reducing device dimensions looking for higher frequencies.  We have demonstrated that we are technologically quite close to this limit. Because of the discrete nature of electrons, noise appears in the electrical total current making impossible to distinguish between different current levels. Only when this current is time averaged we can distinguish between levels, but then the lower operating frequency is not longer related to the transit time.  We remark that disruptive technologies based on innovative working principles, for example, those involving photonic manipulation of information \cite{ref1}, may not be affected by this noise limit and could be used to overcome it. 

As a byproduct of the present work, we also argue that dealing with the linear wave function solution of the Schr\"odinger equation (or its equivalent Dirac equation) is a valid strategy for steady-state quantum electron devices. However, for the simulation of high frequency ultra-small devices, a multi-time measurement process of the particle and displacement current, has to be included in the modeling when looking for noise (time-correlations). It is in this sense that we invoke the need for a second revolution for the electron device simulations to provide the industry with reliable predictions about noise, AC, and transient properties of these new ultra-small quantum electron devices. The BITLLES simulator presented here is an excellent tool to study such high-frequency scenarios in quantum scenarios. It is a great merit of this work to tackle the classical and quantum problem of the THz noise in ultra-small devices with the same fundamental language: electron trajectories. This fact greatly contributes to an easy understanding of the fundamental problem and its important practical consequences.

An application of the BITLLES simulator for a double gate GFET, defining the particle nature of electrons from a Bohmian trajectory and their wave nature from the bispinor solution of the time-dependent Dirac equation, confirms the predictions mentioned above about noise. There are two intrinsic and different limits for determining the maximum working frequency of ultra-small devices. On the one hand, for low frequencies, the transit time limit is the one that established a maximum value for the working frequency. At higher frequencies, due to the device miniaturization and because of the discreteness of the few electrons being in the system, the noise limit cannot longer be neglected and it competes with the transit time limit. At high enough frequencies it can even overcome the transit time limit (see \fref{fig_ana}, where we plotted the ratio between both, the transit time and noise, working frequencies, as well as many data for different state-of-the-art laboratory prototypes and commercial transistors), and then it will represent the true and unavoidable fundamental limitation to reach THz frequencies with ultra-small devices. For instance, new technologies going beyond CMOS  are nowadays completely into the nanoscale regime. For that reason, we predict that some of these new prototypes where channel lengths are around tens of nanometers (such as ferroelectric field effect transistors  \cite{fefet}, carbon nanotubes \cite{cn,cn2,cn3},  nanowires \cite{nw1,nw2,nw3} or other laboratory prototypes \cite{gaat,mos2,grt}) can be completely adequate for DC applications, but will fail when trying to operate at THz frequencies. Such THz noise cannot be avoided in ultra-small devices because it is directly linked to the discrete nature of the few electrons present in the active region. The only way of overcoming this noise is enlarging the dimensions of the active device region to accommodate more electrons inside. But, this solution is contrary to the common scaling strategies for nanoscale devices. Finally, we notice that the plasmons in the contacts (as an additional source of THz noise) have not been considered in this work.  Therefore, the dramatic effect of this unexpected THz noise in limiting the real speed of ultra-small devices can be even worse than what we have predicted \cite{DamianoPRL}.


\section*{Acknowledgments}

We acknowledge funding from the ``Ministerio de Ciencia, Innovaci\'{o}n y Universidades'' under Grant No. RTI2018-097876-B-C21 (MCIU/AEI/FEDER, UE) and TEC2017-83910-R, the Consejer\'{i}a de Educaci\'{o}n de la Junta de Castilla y Le\'{o}n (project SA254P18), the Generalitat de Catalunya and FEDER for the project QUANTUMCAT 001-P-001644, the European Union's Horizon 2020 research and innovation programme under grant agreement No Graphene Core2 785219 and under the Marie Skodowska-Curie grant agreement No 765426 (TeraApps). 

\section*{Appendix}

\subsection{The equation of motion for individual electrons in quantum systems that are being continuously measured}

In high-frequency scenarios, electrons in ultra-samll devices can be considered as quantum system being continuously measured. Most approaches for open quantum systems revolve around the reduced density matrix constructed by tracing out the degrees of freedom of the environment (or measuring apparatus) \cite{open}. For Markovian evolutions, the Lindblad master equation \cite{lindblad} preserves complete positivity \cite{vega}, but its connection to realistic practical scenarios and its extension beyond Markovian dynamics are still challenging \cite{ferialdi,vega}. Alternatively, inspired by spontaneous collapse theories \cite{GRW}, stochastic Schr\"odinger equations (SSEs) \textit{unravel} the reduced density matrix in non-Markovian systems \cite{SSE} in terms of states asigned to a particiular experiment. Continuous measurement theory based on SSE allows the definition of a wave function of the open system conditioned on one \textit{monitored} value associated with the environment (or measuring apparatus) \cite{wiseman,wiseman1,Gambetta,dioosi}. In practical applications, the non-hermitian Hamiltonians that govern such \textit{conditioned} wave function can provoke states of the SSE to lose their norm and therefore their statistical relevance \cite{open}. It has been shown that linking those (conditioned) wave functions at different times assigning them physical reality (beyond mere mathematical elements to properly reproduce ensemble values) requires dealing with theories that allow a well-defiend description of some properties (here the electrical current) even in the absence of measurement\cite{Gambetta,dioosi,wiseman}.

Under the Bohmian theory, we can tackle this problem through the use of the  conditional wave function \cite{Bohm}. Such a conditional wave function provides an unproblematic way of defining the wave function of a subsystem (i.e. open system), either from a computational and an interpretative points of view. By construction, within Bohmian mechanics, the conditional wave function is always a well-defined physical state for Markovian and non-Markovian open systems, with continuous or non-continuous measurements. In this work, we have used this simulation technique when dealing with quantum electron device simulations. Since this approach deals directly with the wave function as a guiding field of the trajectory, it provides a completely positive map for either Markovian or non-Markovian dynamics with an unproblematic physical interpretation of the wave function of the open system at different times (for more details, see \cite{EnriquePRB}). 

\subsubsection{Application to electron devices governed by the Schr\"odinger equation} 
The general expression of the equations of motion of such a conditional wave function is explained in Ref. \cite{OriolsPRL}.  Here we provide a brief summary. Let us consider an \textit{isolated} (\textit{closed}) quantum system described by a full many-body state $|\Psi \rangle$ solution of the unitary, reversible, and linear Schr\"odinger equation. We decompose the total Hilbert space of $N$ particles in two sets, one with the particle under study ($a$ subset) and the other particles ($b$ subset, which includes the apparatus particles) as $\mathcal{\hat H}=\mathcal{\hat H}_a \otimes \mathcal{\hat H}_b$, with $\vec r =\{\vec r_a,\vec r_b\}$ being $\vec r_a$  the position of the $a$-particle and $\vec {\textbf{r}}_b=\{\vec r_1,..,\vec r_{a-1},\vec r_{a+1},.., \vec r_N\}$ the position of all other particles.

It has been shown in Ref. \cite{OriolsPRL} that, for each experiment labeled by $j$, the conditional wave function $\psi_a^j$ can be computed, in general, from the following single-particle Schr\"odinger-like equation in physical space:
\begin{eqnarray}
\label{Pseu_Sch}
i \hbar\frac {d \langle  \vec r | \Psi \rangle }{dt}\Big|_{\vec {\textbf{r}}_b^j[t]}= \langle \vec r | \mathcal{\hat H} |\Psi \rangle|_{\vec {\textbf{r}}_b^j[t]} \Longleftrightarrow i \hbar\frac {d \psi_a^j}{dt}={H_a} \psi_a^j
\end{eqnarray}
where $\mathcal{\hat H}$ is the many-body Hamiltonian and its relation to $H_a$ will be explained next. We define $\vec {\textbf{r}}_a^j[t]$ as the Bohmian trajectory of the $a$-particle and $\vec {\textbf{r}}_b^j[t]$ represents the actual positions of the $b$ particles. Let us notice that the relation between $i \hbar {d \langle  \vec r | \Psi \rangle }/{dt}|_{\vec {\textbf{r}}_b^j[t]}$ and $i \hbar {d \psi_a^j}/{dt}$ on the right and left sides of \eref{Pseu_Sch} is the following:
\begin{eqnarray}
i \hbar\frac {d \psi_a^j(\vec r_a,t)}{dt} &=& i \hbar\frac {d \langle r_a,{\vec {\textbf{r}}_b^j[t]} | \Psi(t) \rangle }{dt}= i \hbar\frac {d \langle \vec r | \Psi(t) \rangle }{dt}\Big|_{\vec {\textbf{r}}_b^j[t]}+i \hbar\sum_{k=1,k\neq a}^{N} \nabla_k  \langle \vec r | \Psi(t) \rangle\Big|_{\vec {\textbf{r}}_b^j[t]}  \vec v_k^j[t] \nonumber\\&=&i \hbar\frac {d \langle \vec r | \Psi(t) \rangle }{dt}\Big|_{\vec {\textbf{r}}_b^j[t]}\!\!+i B_a(\vec r_a,{\vec {\textbf{r}}_b^j[t]},t)
\label{t04}
\end{eqnarray}
with the conditional imaginary potential $iB_a$ defined as:
\begin{eqnarray}
B_a\equiv \hbar\sum_{k=1,k\neq a}^{N} \nabla_k  \langle \vec r | \Psi(t) \rangle\Big|_{\vec {\textbf{r}}_b^j[t]}  \vec v_k^j[t]
\label{B}
\end{eqnarray}
where  $\vec v_k^j[t]={d \vec r_k^j[t]}/{dt}$ is the Bohmian velocity  of the $k$ particle given by
\begin{equation}
\label{velo}
\vec v^j_a[t]=\frac{d \vec r^j_a[t]}{dt}=\frac {\vec J_a(\vec r_a^j[t],\vec {\textbf{r}}_b^j[t],t)}{|\Psi(\vec r_a^j[t],\vec {\textbf{r}}_b^j[t],t)|^2} ,
\end{equation}
where $\vec J_ a = \hbar \mathop{\rm Im} (\Psi^* \nabla_a \Psi)/m_a$ is the (ensemble value of the) current density with $m_a$ the mass of the $a$-th particle. Once we have defined $B_a$, the term $H_a$ on the right hand side of \eref{Pseu_Sch} can be defined as:
\begin{eqnarray}
H_a=\frac {\langle \vec r | \mathcal{\hat H} |\Psi(t) \rangle|_{\vec {\textbf{r}}_b^j[t]}+iB_a} {\psi_a^j}
\label{H_a}
\end{eqnarray}
In general, \eref{Pseu_Sch} is non-linear because $H_a$ in \eref{H_a} depends on the wave function itself. In addition, the imaginary conditional potential $iB_a$ indicates  that the evolution of the CWF can be non-unitary.  \eref{Pseu_Sch} includes any type of evolution for the conditional wave function (not only linear and unitary ones) and, in particular,  it allows the description of irreversible dynamics in  open system when a continuous measurement is performed, as required in this work. Obviously, the full wave function $\Psi(\vec r_a,\vec {\textbf{r}}_{b},t)$ satisfies unitary and linear dynamics, with conservation of the total energy \cite{libro}.

The key computation for the practical application of our approach is the evaluation of $H_a$ in \eref{H_a}, which allows us to determine an equation of motion for each conditional wave function. The calculation of $\langle \vec r | \mathcal{\hat H} |\Psi(t) \rangle$ before conditioning depends on the full many-body wave function and it requires educated guesses \cite{OriolsPRL,libro}. The potential $B_a$, which contains many-body terms but it does not depend directly on $\mathcal{\hat H}$, will be approximated following Ref. \cite{OriolsPRL}. Stochasticity is introduced in \eref{Pseu_Sch} through the term  $H_a$ which accounts for the effect of non-simulated degrees of freedom of the environment in each experiment and from the initial values defining the conditional wave function $\psi_a^j$ and it trajectory  $\vec {\textbf{r}}_a^j[t]$.\\

As said in the text, in the case of the displacement current one of the authors showed recently that the measurement of the displacement current in a quantum system can be considered as a type of weak measurement \cite{DamianoPRL}. For that reason, we consider $B_a\equiv 0$ as a good estimation for the present scenario. In this work, with the BITLLES simulator, we are just computing the autocorrelation (noise) given by the electrons in the quantum system without considering the contribution from the metallic contacts. For this reason, we have argued in the text that our dramatic predictions about the impossibility of ultra-small devices to work at THz devices is developed, in fact, for the best scenario for these ultra-small devices (when the additional noise of the contacts is neglected). The experimental noise results can be worst than what we have predicted here. 

\subsubsection{Application to electron devices governed by the Dirac equation} 
Under the above approximation (weak measurement, i.e., $B_a\equiv 0$), graphene dynamics are just given by the Dirac equation, and not by the usual Schr\"{o}dinger one. The presence of the Dirac equation on the description of the dynamics of electrons in graphene is not due to any relativistic correction, but to the presence of a linear energy-momentum dispersion (in fact, the graphene Fermi velocity $v_f=10^6 \: m/s$ is faster than the electron velocity in typical parabolic band materials, but still some orders of magnitude slower than the speed of light). Thus, the conditional wave function associated to the electron is no longer a scalar, but a bispinor. In particular, the initial bispinor is defined (located outside of the active region) as: 
\begin{equation}
\label{bispinor}
\begin{pmatrix}
\psi_1(x,z,t) \\
\psi_2(x,z,t) 
\end{pmatrix}= \left(\begin{matrix}
1 \\ se^{i\theta_{\vec{k_c}}}
\end{matrix}\right)\Psi_g(x,z,t)
\end{equation}
where $\Psi_g(x,z,t)$ is a gaussian function with central momentum $\vec{k_c}=(k_{x,c},k_{z,c})$, $s=1$ ($s=-1$) if the electron is in the CB (VB) and $\theta_{\vec{k_c}}=atan(k_{z,c}/k_{x,c})$. The wave packet can be considered as a Bohmian conditional wave function for the electron, a unique tool of Bohmian mechanics that allows to tackle the many-body and measurement problems in a computationally efficient way \cite{OriolsPRL,EnriquePRB}. The two components are solution of the mentioned Dirac equation:
\begin{eqnarray}
\label{dirac2d2}
i \hbar \frac{\partial }{\partial t}
\begin{pmatrix}
\psi_1\\ 
\psi_2 
\end{pmatrix}=
\begin{pmatrix}
V(x,z,t) & -i\hbar v_f \frac{\partial }{\partial x}-\hbar v_f \frac{\partial }{\partial z}\\
-i\hbar v_f \frac{\partial }{\partial x}+\hbar v_f \frac{\partial }{\partial z} & V(x,z,t)
\end{pmatrix} 
\begin{pmatrix}
\psi_1 \\
\psi_2 
\end{pmatrix}=
&-i\hbar v_f \left(\vec{\sigma} \cdot \vec{\nabla}+V\right)\left( \begin{array}{lcr}
\psi_1 \\
\psi_2 
      \end{array}
    \right)
\end{eqnarray}
where $v_f=10^6 \: m/s$ is the mentioned Fermi velocity and $V(x,z,t)$ is the electrostatic potential. $\vec{\sigma} $ are the Pauli matrices:
\begin{eqnarray}
\vec{\sigma}=(\sigma_x,\sigma_z)=
\left(
\begin{pmatrix}
0&1\\
1&0\\
\end{pmatrix}
,
\begin{pmatrix}
0&-i\\
i&0\\
\end{pmatrix}
\right)
\label{11}
\end{eqnarray}
Usually, in the literature, one finds $\sigma_z$ as $\sigma_y$, however, since we defined the graphene plane as the $XZ$ one, the notation here is different. From \eref{11} we can obtain a continuity equation for the Dirac equation and then we can easily identify the Bohmian velocities of electrons as \cite{libro}
\begin{equation}
\vec{v}(\vec{r},t)=\dfrac{ v_f\psi(\vec{r},t)^{\dagger}\vec{\sigma}\psi(\vec{r},t)}{|\psi(\vec{r},t)|^2}
\label{bvel}
\end{equation}
By time integrating \eref{bvel} we can obtain the quantum Bohmian trajectories. The initial positions of the trajectories must be distributed according to the modulus square of the initial wave function, i.e., satisfying the quantum equilibrium hypothesis and thus certifying the same empirical results for ensemble values as the orthodox theory \cite{libro,nino1}. All this formalism was introduced in the BITLLES simulator in order to correctly model graphene and other linear band structure materials. Once the quantum trajectory of the $a$-electron is defined the computation of the electrical current (the correlations and its contribution to the fluctuations) is done exactly as one routinely does for semi-classical simulations. It is a great merit of this work to tackle the classical and quantum problem of the THz noise in ultra-small devices with the same fundamental language: electron trajectories. This fact greatly contributes to an easy understanding of the fundamental problem and its important practical consequences.


\begin{thebibliography}{100}
\bibitem{ft1}
The maximum clock frequency of digital circuits is usually limited to $1/3$ of the experimental cutoff frequency $f_{t}$ of the discrete transistors  including parasitic effects.
\bibitem{ft2}
In this work, the electrical field is always treated as a classical field solution of the Maxwell equations, without photon quantization. This is a typical assumption for high frequency modeling of electron devices. It is based on the facts that (i) the energies (frequency) of the photons at THz frequencies are still small in comparison with the energy of electronic transitions and (ii) the electric fields in nanoelectronic devices is high enough to ignore the quantization in the number of photons.
\bibitem{ft3}
Let us notice that, strictly speaking, the device for $T=\tau$ does not belong to the scenario b. However, we have used the particular value $N_{cross,\tau}$ only to compute the parameter $B$ in \eref{noisb2}. The very same \eref{snr2} will be obtained by considering a $T'=\alpha \tau$ with $\alpha\gg 1$, using $\langle N_{cross,T'} \rangle=\alpha \langle N_{cross,\tau} \rangle$ and $var( N_{cross,T'})=\alpha var (N_{cross,\tau})$ showing that the exact value of $\alpha$ is irrelevant for computing $B$. We chose $T=\tau$ in order to be able of comparing the transit time limit with the noise limit.
\end{thebibliography}
\end{document}